\documentclass{aa}
\usepackage{graphics}

\begin{document}

%\thesaurus{08. (08.22.3)         % Stars: variables: other
%     11.              % A&A Section 11: Galaxies
%    (11.12.1;             % Galaxies: Local Group
%    11.09.1 IC\,1613;    % Galaxies: individual: IC\,1613
%    11.19.5)             % Galaxies: stellar content
%        }
%
%

\title {The First Known Mira-type Variable Star in IC 1613}

\author{
   R.~Kurtev\inst{1}
   \and
   L.~Georgiev\inst{2}
     \and
   J.~Borissova\inst{3}
     \and
   W.~D.~Li\inst{4}
     \and
 A.~V.~Filippenko\inst{4}
     \and
 R.~R.~Treffers\inst{4}
}

\offprints{R.~Kurtev}

\institute {Department of Astronomy, Sofia University, and Isaac Newton
Institute of Chile, Bulgarian Branch, BG\,--\,1164 Sofia, Bulgaria \\
email: kurtev@phys.uni-sofia.bg
\and
Instituto de Astronom\'{\i}a, Universidad Nacional Aut\'onoma
   de M\'exico, M\'exico \\
 email: georgiev@astroscu.unam.mx
\and
Institute of Astronomy, Bulgarian Academy of Sciences, and Isaac Newton
Institute of Chile, Bulgarian Branch,
   72~Tsarigradsko Chauss\`ee, BG\,--\,1784 Sofia, Bulgaria \\
email: jura@haemimont.bg
\and
Department of Astronomy, University of California, Berkeley, CA
94720\,--\,3411
USA\\
 email: (wli, alex, rtreffers)@astro.berkeley.edu
}

\date{Received  ............... ; accepted ............... }

\authorrunning {Kurtev et al.}
\titlerunning {Mira-type Variable in IC~1613}

\abstract{
King, Modjaz, \& Li (1999) discovered Nova 1999 in IC~1613 at Lick
Observatory.
Both Fugazza et al. (2000) and Borissova et al. (2000) questioned this
classification,
because they were able to detect the star on images obtained in previous
years.
In infrared frames taken on Oct. 15, 1998, the Nova 1999 has $(J-K) = 1.14$
and
$K = 14.69$ mag. Our light curve study, based primarily on 92 unfiltered
Lick images,
suggested that the object could be a Mira-type variable with a period of
$640.7$ days.
This period is very close to that obtained by Fugazza et al. (2000) ---
$631$ days.
The star is overluminous with respect to the period-luminosity (PL)
relation derived by Feast et al. (1989) for Mira variables in the LMC. At
longer periods $(P > 400~ \rm days)$, many LMC Miras show such behavior and
the
PL relation appears to break down. It is possible that the situation in
IC~1613
is similar. An optical spectrum obtained with the Keck-II telescope shows
features typical of M3Ie or M3IIIe stars. We conclude that the star is a
normal
long-period M-type Mira variable, the first such star confirmed in IC~1613.
}

\maketitle

\section{Introduction}

IC~1613 is a faint irregular member of the Local Group. The galaxy was
discovered by Wolf  (1906) with the 16-inch Bruce refractor at
Heidelberg. Baade (1928), using plates taken with the 16-inch Bergedorf
reflector, classified it as a Magellanic-Cloud-type galaxy.

The variable stars in IC~1613 have been investigated by Sandage (1971),
Carlson
\& Sandage (1990), and Freedman (1988a). Eight Wolf-Rayet stars are
suspected
(Armandroff \& Massey 1985) and one of them (WO3) is surrounded by nebulae.
The light curves of Cepheids and other variable stars in Field A of IC~1613,
obtained with CCD unfiltered photometry, have been analyzed by Antonello et
al. (1999).  Later Antonello et al. (2000) and Mantegazza et al. (2001)
observed another three fields in the galaxy (Fields B, C, and D) to detect
short-period Cepheids and to obtain good light curves for Fourier
decomposition. In all four fields (A-D) they found a total of 
more than 300 variable stars
of different types. Very recently Dolphin et al. (2001) presented {\it
Hubble
Space Telescope} WFPC2 $VI$ photometry of a field in the halo of IC~1613 and
found 13 RR Lyrae stars and 11 Cepheids. Using different distance indicators
they found the best distance modulus of IC~1613 to be $\mu_0=24.31 \pm
0.06$.

King, Modjaz, \& Li (1999) discovered Nova 1999 in IC~1613 with the 0.8-m
Katzman Automatic Imaging Telescope (KAIT; Li et al. 2000; Filippenko et
al. 2001) at Lick Observatory. The new object is located at $\alpha(2000) =
1^{\rm h}04^{\rm m}48\fs97$, $\delta(2000) = +2^{\circ}05'28\farcs8$.  In
the
infrared (IR) frames taken on 1998 October 15 at the National Mexican
Astronomical Observatory ``San Pedro Martir,'' Nova 1999 has $(J-K)=
1.14$
and $K=14.69$ mag (Borissova et al. 2000). Its presence in the images taken
one
year before the discovery and its low observed brightness during outburst
call
into question its classification as a nova. Fugazza et al. (2000) also noted
that the nova falls in their Field B (Antonello et al. 2000), and its
coordinates are coincident within the uncertainties with those of their
variable V2950B. This star was observed four years before the announcement
of
the detection as a nova. Fugazza et al. (2000) found a period of 631 days
for
optical variations. These authors also noted a $BVRI$ measurement of this
star
obtained by Freedman (1988b), dating back to 1984.

\section{Observations and Data Reduction}

\subsection{Photometry}

Unfiltered photometry of Nova 1999 was obtained during the interval 1998
July
25 -- 2000 January 4 ($\sim 90$ CCD frames) with KAIT, as a byproduct of
monitoring IC 1613 for supernovae.  The images were obtained with
an
Apogee $512 \times 512$ pixel AP7 CCD camera. The scale at the Cassegrain
focus
of KAIT is $0.8\arcsec$ pixel$^{-1}$ and the total field of view is
$6\farcm8
\times 6\farcm8$.

Additional CCD $BVR$ frames were taken on 1999 September 16, November 3,
December 8, 2000 August 4, and 2001 July 15 with the 2-m Ritchey-Chreti\'en
(RC) telescope of
the Bulgarian National Astronomical Observatory Rozhen with a Photometrics
$1024 \times 1024$ pixel CCD camera.  The scale at the Cassegrain focus is
$0.33''$ pixel$^{-1}$ and the total field of view is $5\farcm6 \times
5\farcm6$. Stellar photometry of the frames taken with the 2-m telescope was
performed with the point-spread function fitting
routine ALLSTAR available in DAOPHOT II (Stetson 1993). Reduction of
instrumental magnitudes to the standard $BVR$ system was accomplished using
a
few dozen observations of Landolt (1992) standard stars on each night.

Instrumental stellar magnitudes from the unfiltered KAIT observations were
also
derived with ALLSTAR available in DAOPHOT II (Stetson 1993).
Using standard $R$ magnitudes and $V-R$ colors of the variable star and
twelve local standard stars in the filed obtained from Rozhen $BVR$
photometry,
KAIT unfiltered instrumental magnitudes were transformed into Cousins $R$.
A detailed discussion of the procedure can be found in Riess et al. (1999).
The precision of the KAIT observations as a function
of magnitude is given in Table~\ref{tab1}.

\begin{table}[h]
\begin{center}
\caption{Uncertainty of KAIT photometry.}
\tabcolsep=12pt
\begin {tabular}{r r}
\hline
$R$     & $\sigma (R)$ \\
\hline
18.0 & 0.12 \\
18.5 & 0.13 \\
19.0 & 0.17 \\
19.5 & 0.23 \\
20.0 & 0.33 \\
20.5 & 0.52 \\
\hline
\end{tabular}
\label{tab1}
\end{center}
\end{table}

The IR data were acquired with the infrared camera ``CAMILA" with a NICMOS3
$256 \times 256$ pixel detector attached to the 2.1-m telescope of the
National
Mexican Observatory.  The scale is $0\farcs85$ pixel$^{-1}$, resulting in a
field
size of about $3\farcm6 \times 3\farcm6$. A set of $JK$ frames was taken on
1998 October 13--15. The seeing during these observations was $1'' -
1\farcs2$,
with stable photometric conditions. Stellar photometry of the
frames was done using DAOPHOT II (Stetson 1993). Twelve UKIRT standard stars
(Cassaly \& Hawarden 1992) were measured before and after the nova
observations.

\subsection{Spectroscopy}

Spectroscopic observations were made with the Low Resolution Imaging
Spectrometer (LRIS; Oke et al. 1995) on the Keck-II 10-m telescope. Two
spectra
of the program star were obtained in photometric conditions on 1999 December
15. The seeing varied from $1''$ to $1\farcs4$, and the slit width was
$1''$. A
300 grooves mm$^{-1}$ grating blazed at 5000~\AA\ gave a wavelength range of
$3900-8900$~\AA\ (2.5~\AA\ pixel$^{-1}$), with minimal second-order
contamination beyond 7600~\AA. The spectral resolution was 9.9~\AA\ (FWHM of
the [O~I] $\lambda$6300 night-sky line).  The exposure time for each
spectrum
was 300~s. Both spectra were obtained at the parallactic angle of
$130^\circ$
(Filippenko 1982) at an airmass of 1.11.

The spectra were flux-calibrated using the standard star BD+17$^\circ$4708
(Oke
\& Gunn 1983) at airmass 1.06.  The flux standard was also used to remove
telluric absorption lines through appropriate division.  The two exposures
were
combined equally after scaling the first by a factor of 1.07.  
Cosmic rays and/or sky-line
glitches were removed in the regions $6552.5-6660$~\AA, $4097.2-4100$~\AA,
and
4470~\AA\ (a single pixel).  From our final merged spectrum, we derive $VRI$
magnitudes of the variable star.

%                                     Two column figure (place early!)
%______________________________________________Fig.1
   \begin{figure}[h]
   \centering
    \resizebox{\hsize}{!}{\includegraphics{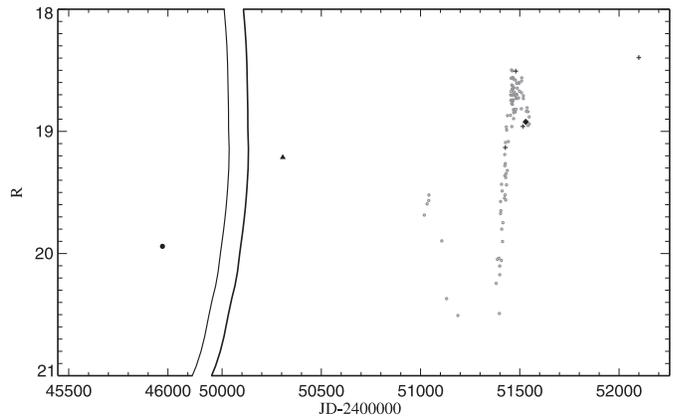}}
   \caption{The long-term photometric behavior of the Mira-type variable in
IC~1613
(``Nova 1999''). Gray circles are KAIT observations transformed to $R$,
crosses 
represent Rozhen observations, and the filled diamond is the $R$ magnitude
derived
from the Keck-II spectrum. The Freedman (1988b) magnitude of
19.94 on JD~2,445,973.88 (filled dot) and the Fugazza et al. (2000) magnitude
of 19.21 on JD~2,450,305.90 (filled triangle) are also showh.
}
              \label{fig1}
    \end{figure}

\section{Photometric Behavior}

Figure~\ref{fig1} shows the $R$-band light curve of Nova 1999, using the observations
of
Freedman (1988b), Fugazza et al. (2000), the Rozhen 2-m RC telescope, the
transformed KAIT data, and the $R$ magnitude derived from the Keck
spectrum. Initially the star was suspected to be a nova in IC~1613 by King
et
al. (1999). Its long-term photometric behavior, however, is typical of
Mira-type variable stars.

\begin{figure}
   \centering
    \resizebox{\hsize}{!}{\includegraphics{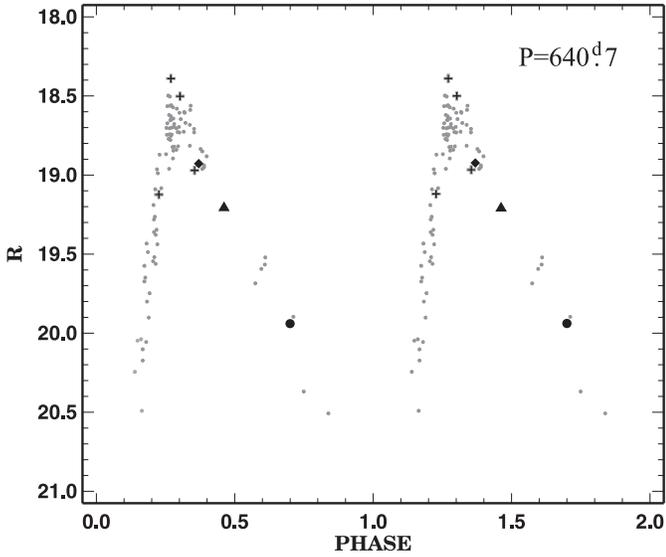}}
   \caption{The mean $R$-band light curve of the Mira-type variable in
IC~1613. The
period is $640.7$ days. The symbols are identical to those in
Figure~\ref{fig1}.
}
              \label{fig2}
    \end{figure}

We searched for the best period using a least-squares periodogram analysis
by
means of the phase dispersion minimization (PDM) task available in IRAF, and
with a period-finding program based on the Lafler \& Kinman (1965) ``theta"
statistic. The result is $640.7$ days. This value is near the upper limit
for Miras. The mean light curve from all observations is displayed in 
Figure~\ref{fig2}. The amplitude of the star is more than $2.5$ mag in $R$, also typical for
this
type of red variable.

%                                     Two column figure (place early!)
%______________________________________________Fig.2

The infrared photometry gives $(J-K) = 1.14$ and $K = 14.69$ mag on
JD~2,451,100. As can be easily seen from Figure~\ref{fig1}, at this moment the $K$
magnitude of the star is somewhat less but close to its average $K$
magnitude. Taking into
account that the amplitudes of Mira-type variables are minimal in the
infrared
(not more than 1 mag), it is interesting to check whether this object lies
on
the Mira period-luminosity (PL) relation found by Feast et al. (1989) in the
LMC: $M_K = -3.47\log (P/{\rm day}) + 0.78$.  
According to this equation, the expected
$M_K$ is $-8.96$ mag. For comparison, if we use the average observed $K$
magnitude
and an adopted distance modulus of $24.31 \pm 0.06$ mag (Dolphin et al. 2001),
we
obtain $M_K = -9.62$ mag. Thus, the star appears to be overluminous
by about 0.7 mag with respect
to
the PL relation derived by Feast et al. (1989) for LMC Miras.
At long periods
$(P > 400\ \rm days)$, many LMC Miras show such behavior and the relation
appears to break down. It is possible that the situation in IC~1613 is
similar.

\section{Spectral Classification}

\begin{figure}
   \centering
    \resizebox{\hsize}{!}{\includegraphics{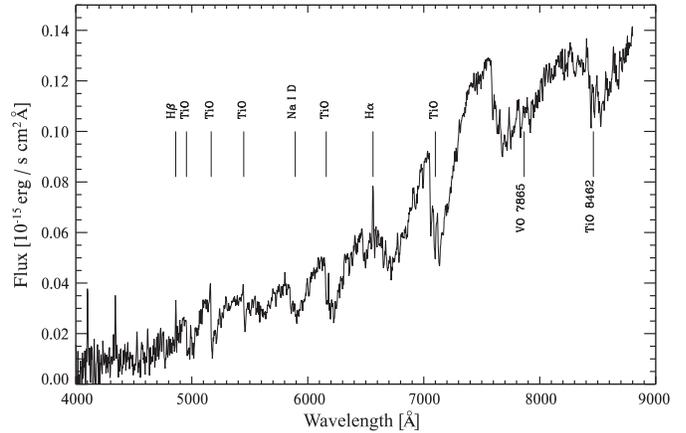}}
   \caption{Keck-II optical spectrum of the Mira variable star.
}
              \label{fig3}
    \end{figure}

The spectra of very red stars are dominated by bands of $\rm TiO$ and other
molecules, including VO. Figure~\ref{fig3} shows the spectrum of the 1999 Nova with
the
main spectral features marked.  The depth of the bands is proportional to
the
temperature (spectral class) of the star. To quantify this relation
O'Connell
(1973) introduced spectral indexes defined as

%                                     Two column figure (place early!)
%______________________________________________Fig.3

\begin{eqnarray*}
[\rm TiO]_1 &=& -2.5 \log\big(F_{6180}/[F_{6125}+(F_{6370}- F_{6125}) \\
&& (6180-6125)/(6370-6125)]\big),
\end{eqnarray*}

\begin{eqnarray*}
[\rm TiO]_2 &=& -2.5 \log\big(F_{7100}/[F_{7025}+(F_{7400}-F_{7025}) \\
&&(7100-7025)/(7400-7025)]\big),
\end{eqnarray*}

\begin{eqnarray*}
[\rm VO] &=& -2.5 \log\big(F_{7865}/[F_{7400}+(F_{8050}-F_{7400}) \\
&&(8050-7865)/(8050-7400)]\big).
\end{eqnarray*}

\noindent
The fluxes are measured in 30~\AA\ bandpasses centered at the corresponding
wavelengths.  The positions of the bandpasses for the two $\rm TiO$ indexes
are
shown in Figures~4a and 4b.  Based on several standard stars, Kenyon \&
Fernandez-Castro (1987) obtained a calibration of these indexes in the K and
M
spectral classes. Similar calibration but for the $\rm TiO$ band
at 8462~\AA\ was used by Zhu et al. (1999). We will call this index $[\rm
TiO]_3$.

%                                     Two column figure (place early!)
%______________________________________________Fig.4
   \begin{figure*}
   \centering
    \resizebox{\hsize}{!}{\includegraphics{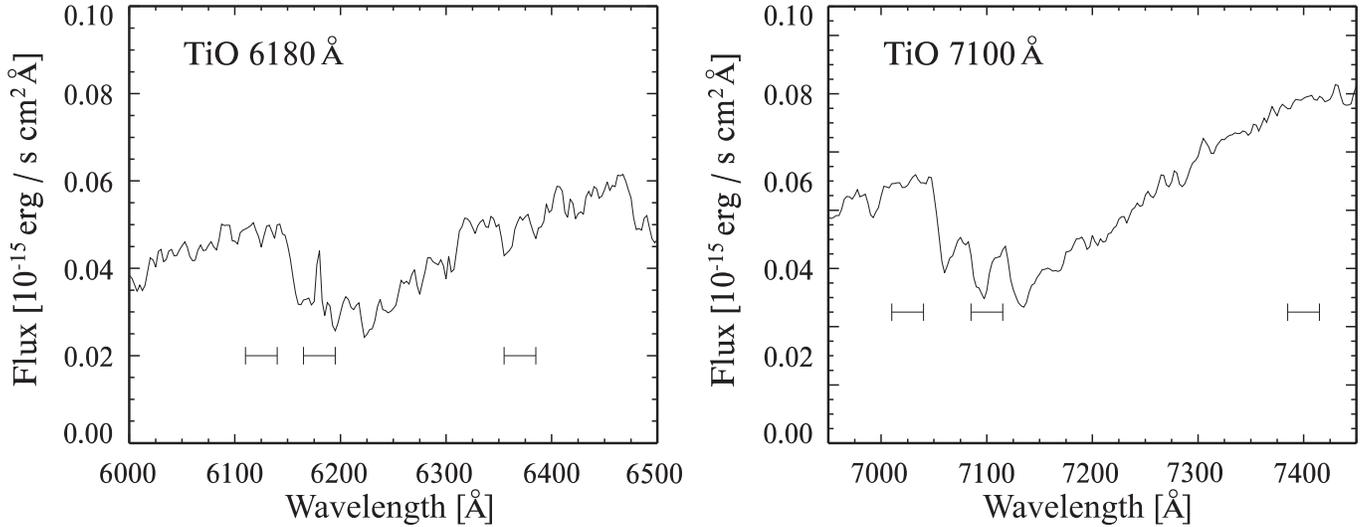}}
   \caption{TiO molecular bands at (a) 6180~\AA\ and (b) 7100~\AA. The bars
show
the intervals used to construct the indexes.
}
              \label{fig4}
    \end{figure*}

To determine the spectral type of the star we first corrected the spectrum
for
the radial velocity of IC~1613, $-238$ km s$^{-1}$. Then we measured the
flux
at the bandpasses for each index. The results are given in Table~\ref{tab2},
together
with the quantity ST [see equations 5, 6, and 7 of Kenyon \&
Fernandez-Castro
(1987) and equation 1 of Zhu et al. (1999)] and the corresponding spectral
class.

\begin{table}
\begin{center}
\caption{Spectral indexes and classification.}
\tabcolsep=2.3pt
\begin {tabular}{l c c c}
\hline
Index     & Value   &  $ST$ & Sp. Class \\
\hline
$[\rm TiO]_1$ & 0.43    &   2.3 &   M2      \\
$[\rm TiO]_2$ & 0.53    &   2.7 &   M3      \\
$[\rm TiO]_3$ & 0.13    &   3.3 &   M3      \\
$[\rm VO]$    & 0.13    &   3.0 &   M3      \\
$[\rm Ca~II]$ & 0.20    &       &           \\
\hline
\end{tabular}
\label{tab2}
\end{center}
\end{table}

The relatively early spectral type given by the $[\rm TiO]_1$ index is due
to
an ``emission" feature at 6180~\AA\ which falls in the bandpass and
decreases
the index. The other three indexes give a consistent spectral type, M3. The
determination of the luminosity class is much more uncertain.  All of the
proposed indicators are applicable to an M3 star, or the resolution of our
spectrum is not sufficiently high to measure them. The only measurable index
sensitive to the luminosity is Ca~II $\lambda$8542 (O'Connell 1973). The
value
of the Ca~II index plotted on Figure~7 of O'Connell (1973) falls in the
region
of giants and supergiants.  It is not possible to distinguish between the
M3I and
M3III luminosity classes.  Borissova et al. (2000) estimated the bolometric
magnitude of the star to be $-6.65$, based on $JK$ photometry,
suggesting luminosity class I. The star also shows hydrogen Balmer lines in
emission, so the spectral class should be M3Ie or M3IIIe.

\section{Is the Star a Mira?}

The red variables are classified as Miras, semiregular, and slow irregular
variables.  Our light curve covers only one maximum well, and the
classification of the star into one of these groups based on the photometry
alone is not obvious. We know that the star has a period of $640.7$ days.
The $R$ amplitude of the star is around $2.5 - 3$ mag and the light curve is
very
asymmetric (Figure~\ref{fig2}).  Following Mattei et al. (1997), we could reject the
semiregular and slow irregular classifications because of the large
amplitude
of the variations. Plotting the star on Figure~2 of Mattei et al. (1997), it
falls in the region occupied by M-type and S-type Miras.  Mennessier et
al. (1997) classified the red variables in our Galaxy based on several
criteria. 

%                                     Two column figure (place early!)
%______________________________________________Fig.5
   \begin{figure}[h]
   \centering
    \resizebox{\hsize}{!}{\includegraphics{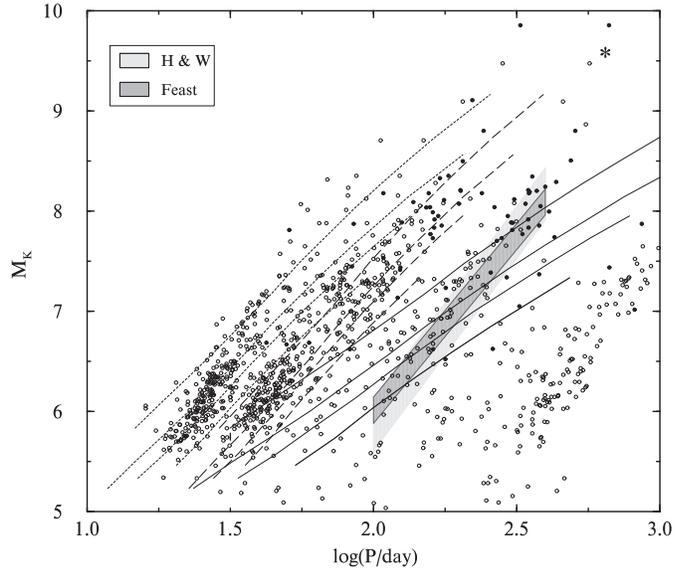}}
   \caption{This is Figure 6 of Barthes \& Luri (2001). It represents the PL
distribution of red variable stars of the LMC in the MACHO data base. The
magnitudes in this figure are taken from Wood (1999).  The calibrated models
are given: fundamental mode (solid lines), first overtone (long-dashed
lines),
and second overtone (dashed lines). The Mira-like PL relations of Feast et
al. (1989) and Hughes \& Wood (1990; H \& W) are also shown. The
investigated
variable star (asterisk) falls in the zone of the first-overtone pulsating
Miras in the LMC.
}
              \label{fig5}
    \end{figure}

\noindent We have three of them: the amplitude, the period, and the
asymmetry
of the light curve. The large amplitude and the long period again put the
star
between M and S-type Miras. But the light curve is quite asymmetric (the
rising
part is about 30\% of the period); hence, the star is most probably an
M-type
Mira.

As mentioned above, the bolometric magnitude of the star is 
$-6.65$.  The PL relation for Miras (Hughes \& Wood 1990) is

$$\langle M_{\rm bol}\rangle = -3.22 - 7.76 [\log (P/{\rm day}) 
- 2.4] \pm 0.38,$$

\noindent
from which we predict a bolometric magnitude of $-6.37 \pm 0.38$. Including
also uncertainties in the period and photometric magnitude, 
the two estimates are completely consistent.

  In Figure~\ref{fig5}, we plot the star on Figure 6 of Barthes \& Luri (2001)
representing the PL distribution of LMC red variable stars in the MACHO data
base.  Observational data are compared to the calibrated models, and the
Mira-like PL relations of Feast et al. (1989) and Hughes \& Wood (1990) are
also shown. As can be easily seen, the variable star (asterisk in Figure~\ref{fig5})
falls
in the zone of the first-overtone pulsating LMC Miras.

We can conclude that ``Nova 1999'' is a normal long-period M-type Mira
variable
star. This is the first confirmed Mira star in IC~1613. The similarity
between
this star and the LMC Miras suggests that the metallicity does not play an
important role in the physics of the pulsation of the Miras (Feast 1996).

\begin{acknowledgements}

The authors gratefully acknowledge  the useful comments and suggestions
made by referees Drs. E. Antonello and L. Mantegazza.
The W. M. Keck Observatory is operated as a scientific partnership between
Caltech, the University of California, and NASA; it was funded by the
generous
contributions of the W. M. Keck Foundation.  We thank Tom Matheson for
calibrating the Keck spectrum. Financial support for this work was provided
to
A.V.F. by NSF grant AST-9987438 and by the Guggenheim Foundation. KAIT was
made
possible by donations from Sun Microsystems Inc., the Hewlett-Packard
Company,
AutoScope Corporation, Lick Observatory, the US National Science Foundation,
the University of California, and the Sylvia and Jim Katzman Foundation.
This
research was supported in part by the Bulgarian National Science Foundation
grant under contract No. F-812/1998 with the Bulgarian Ministry of Education
and Sciences.

\end{acknowledgements}

\newpage

\end{document}